\documentclass[a4paper,11pt]{article}
\usepackage{pos}
\usepackage{placeins}

\title{Taylor expansions and Pad\'e approximations for Lefschetz thimbles and beyond}

\author*{Kevin Zambello}
\author{Francesco Di Renzo}
\author{Simran Singh}

\affiliation{Dipartimento di Scienze Matematiche, Fisiche e Informatiche, Universit\`a di Parma and INFN, Gruppo Collegato di Parma, I-43124 Parma, Italy}

\emailAdd{kevin.zambello@pr.infn.it}
\emailAdd{francesco.direnzo@pr.infn.it}
\emailAdd{simran.singh@unipr.it}

\abstract{Deforming the domain of integration after complexification of the field variables is an intriguing idea to tackle the sign problem.
In thimble regularization the domain of integration is deformed into an union of manifolds called Lefschetz thimbles.
On each thimble the imaginary part of the action stays constant and the sign problem disappears. A long standing issue of this approach is how
to determine the relative weight to assign to each thimble contribution in the (multi)-thimble decomposition. Yet this is an issue one has to face,
as previous work has shown that different theories exist for which the contributions coming from thimbles other than the dominant one cannot be
neglected. Historically, one of the first examples of such theories is the one-dimensional Thirring model.
Here we discuss how Taylor expansions can be used to by-pass the need for multi-thimble simulations. If multiple, disjoint regions can be found
in the parameters space of the theory where only one thimble gives a relevant contribution, multiple Taylor expansions can be carried out in
those regions to reach other regions by single thimble simulations. Better yet, these Taylor expansions can be bridged by Padé interpolants.
Not only does this improve the convergence properties of the series, but it also gives access to information about the analytical structure
of the observables. The true singularities of the observables can be recovered. We show that this program can be applied to the
one-dimensional Thirring model and to a (simple) version of HDQCD. But the general idea behind our strategy can be helpful beyond thimble
regularization itself, i.e. it could be valuable in studying the singularities of QCD in the complex $\mu_B$ plane.
Indeed this is a program that is currently being carried out by the Bielefeld-Parma collaboration.}

\FullConference{%
 The 38th International Symposium on Lattice Field Theory, LATTICE2021
  26th-30th July, 2021
  Zoom/Gather@Massachusetts Institute of Technology
}

\begin{document}
\maketitle

\section{Thimble regularization}
The sign problem is the main obstacle to study quantum field theories (such as QCD) at finite density by lattice simulations.
On the lattice the physical observables have the form
$$\langle O \rangle = \frac{1}{Z} \int dx ~\langle O(x) \rangle e^{-S(x)} = \frac{\int dx ~ O(x) e^{-S(x)}}{\int dx ~ e^{-S(x)}}\mbox{ , }$$
where $Z$ is the partition function and $S(x)$ is the action of the system. Assuming $S(x)$ is real,
the Boltzmannian factor $e^{-S(x)}$ is a well-defined probability distribution for importance sampling and the
integrals can be calculated by Monte Carlo. Unfortunately at finite density the action is in general complex-valued.

An intriguing solution to sidestep the sign problem is to complexify the degrees of freedom of the theory and
deform the contour of integration of the integrals in such a way that the sign problem disappears (or is mitigated).
Thimble regularization provides a way to find a suitable deformation \cite{Cristoforetti:2012su,Fujii:2013sra}.
After complexifying the degrees of freedom,
$x \mapsto z$, one looks for the stationary points of the action $S(z)$, i.e. the points $z_\sigma$ such that
$\partial_z S|_{z_\sigma} = 0$. For each critical point $z_\sigma$ one can define a manifold, the (\textit{stable})
Lefschetz thimble $\mathcal{J}_\sigma$, as the set of the solutions of the steepest ascent (SA) equations originating
from the critical points, i.e. the set of all paths $z(t)$ such that $\frac{dz_i}{dt} = \frac{\partial \overline{S}}{\partial \overline{z}_i}$
and $z_i(-\infty) = z_{i,\sigma}$. Due to the properties of the SA equations the imaginary part of the action stays
constant on each thimble and the real part of the action is always increasing.

According to the Picard-Lefschetz theory,
the original integrals can be decomposed to a sum of integrals over the thimbles,
$$\langle O \rangle = \frac{\int dx ~ O(x) e^{-S(x)}}{\int dx ~e^{-S(x)}} = \frac{\sum_\sigma n_\sigma e^{-iS_I^\sigma}  \int_{\mathcal{J}_\sigma} dz~O(z) ~e^{i\omega} e^{-S_R(z)}}{\sum_\sigma n_\sigma e^{-iS_I^\sigma}  \int_{\mathcal{J}_\sigma} dz ~ e^{i \omega} e^{-S_R(z)} }$$
In place the original integrals we have linear combinations of integrals over the thimbles and the latter are not affected by the sign
problem. The reason is that the contribution from the imaginary part of the action is constant and can be factorized out.
A complex \textit{residual} phase, $e^{i\omega}$, still appears in the integrals. This phase comes from the orientation of the thimble
in the embedding manifold. This results in a \textit{residual} sign problem, but in practice this is usually found to be a mild one
and can be taken care of by reweighting. The integer coefficients $n_\sigma$, known as intersection numbers, can be zero, therefore not all the thimbles
do necessarily contribute. Geometrically the intersection numbers count the number of intersections between the original domain of integration and the
\textit{unstable} thimbles. These are defined as the solutions of the steepest descent (SD) equations originating from the critical points.
The imaginary part of the action is constant on the \textit{unstable} thimbles and the real part of the action is always decreasing.
Now if we define the thimble contribution
$$ \langle \bullet \rangle_\sigma = \frac{\int_{\mathcal{J}_\sigma} dz~ \bullet ~e^{-S_R} }{\int_{\mathcal{J}_\sigma}  dz~e^{-S_R} } = \frac{\int_{\mathcal{J}_\sigma} dz~ \bullet ~e^{-S_R} }{ Z_\sigma }$$
we can rewrite the thimble decomposition formula as
$$\frac{\int dx ~ O(x) e^{-S(x)}}{\int dx ~e^{-S(x)}} = \frac{\sum_\sigma n_\sigma e^{-iS_I^\sigma} Z_\sigma \langle O e^{i\omega} \rangle_\sigma }{\sum_\sigma n_\sigma e^{-iS_I^\sigma}  Z_\sigma \langle e^{i\omega} \rangle_\sigma}$$
The original integrals are now decomposed to a sum of thimble contributions $\langle \bullet \rangle_\sigma$ that can be calculated
numerically by Monte Carlo simulations. These contributions are weighted by the partition function $Z_\sigma$.
Calculating these weights is hard and this could be regarded as one of the main obstacles to multi-thimble calculations. 
Yet this is an issue one has to face, as previous work has shown that different theories exist for which the contributions
coming from thimbles other than the dominant one cannot be neglected \cite{ThirringKiku,StudyThirringKiku,QCD01,Zambello:2018ibq,Savvas}.
Some proposals to address the issue have been made in the literature \cite{QCD01, Zambello:2018ibq, reweightHeidelberg, Bielefeld}. Here
we discuss what we think is a more promising approach where the need to calculate the weights is actually bypassed.

\section{Taylor expansions and Pad\'e approximants}
The idea we propose is to take advantage of the changes the thimble structure is subject to as we move within the parameter space of the theory \cite{DiRenzo:2020cgp}.
The union of the stable thimbles appearing in the thimble decomposition is a deformation of the original contour of integration. 
The thimble structure changes as we modify the parameters of the theory. Being solutions of the same differential equations starting from different
initial conditions, the stable thimbles cannot cross each other. Therefore the stable thimbles appearing in the decomposition
act as barriers, preventing the other stable thimbles from entering the decomposition. When a Stokes phenomenon takes place, however, the stable thimble attached
to a critical point overlaps the unstable thimble attached to another critical point. 
After a Stokes phenomenon, a new (old) stable thimble can enter (leave) the decomposition and its intersection number suddenly changes.
There is a discontinuity in the thimble decomposition. However this does not imply that the observables themselves are discontinuous.
Hence if only one \textit{relevant}\footnote{Here we must consider two possible cases: (1) only the \textit{dominant} thimble $\mathcal{J}_{\sigma_0}$
has a non-zero intersection number and (2) also other thimbles have a non-zero intersection number
but their contribution is exponentially suppressed because $S_R(\sigma_i) \gg S_R(\sigma_0)$.}
thimble is left in the decomposition after a Stokes phenomenon takes place at $\mu_c$, we can think of reaching the region
$\mu < \mu_c$ before the Stokes by a Taylor expansion calculated around a point $\mu_0 > \mu_c $ after the Stokes,
$$\langle O \rangle (\mu) = \sum_{n=0}^\infty \frac{1}{n!} \frac{\partial^n \langle O \rangle}{\partial \mu^n}|_{\mu_0} (\mu - \mu_0)^n$$
The Taylor coefficients can be calculated from one-thimble simulations, bypassing de facto the necessity of calculating the weights of the thimbles.

If multiple suitable expansion points $\mu_0^{(k)}$ are available, the Taylor coefficients can also be used to build a multi-point Pad\'e approximant.
That is we look for a rational function $R_{n,m} (\mu) = p_n(\mu) / q_m (\mu)$ that matches the Taylor coefficients, i.e.
$(\partial^j_{\mu^j} R_{n,m})(\mu_0^{(k)}) = (\partial^j_{\mu^j} \langle O \rangle)(\mu_0^{(k)})$. These Pad\'e
approximants improve the convergence properties of the series and allow to extract information on the singularities structure.
The location of the true singularities of the observable in the complex $\mu$ plane can be inferred from the location of the uncancelled
roots of the denominator.

\section{Applications}
As a first application we have applied our strategy to the one-dimensional Thirring model \cite{ThirringKiku,StudyThirringKiku,Alexandru:2015xva} with parameters
$L=8$, $\beta = 1$, $m=2$. We have calculated Taylor expansions in term of the dimensionless parameter $\frac{\mu}{m}$.
As expansion points we have selected four points where only one thimble gives a relevant contribution, i.e.
$\frac{\mu}{m} = 0.0, 0.4, 1.4$ and $1.8$. Up to some value $\frac{\mu_0}{m} > 0.4$ only two unstable thimbles have an imaginary
part of the action in the range taken by $S_I$ on the original domain of integration. These are the thimbles
attached to two critical points that we denote by $\sigma_0$ and $\sigma_{\bar{0}}$. The contribution from the second critical
point is depressed, as $S_R(\sigma_{\bar{0}}) \gg S_R(\sigma_0)$. Therefore only $\sigma_0$  gives a relevant contribution 
at $\mu = 0.0, 0.4$. On the other hand at $\frac{\mu}{m} = 1.4, 1.8$ all critical points but three have a real part of the action
much greater than $S_R(\sigma_0)$ and their contributions are depressed. Two out of the three critical points have a real part of the action
lower than the minimum taken on the original domain of integration, hence they cannot enter the decomposition as
their unstable thimbles cannot intersect the original domain of integration. It can explicitly be checked that the unstable thimble attached to
the remaining critical point, which we denote by $\sigma_1$, does not intersect the original domain of integration,
while the unstable thimble attached to $\sigma_0$ does \cite{DiRenzo:2020cgp}. Therefore $\sigma_0$ gives the only relevant contribution also at $\frac{\mu}{m} = 1.4, 1.8$.

\begin{figure}[htp]
    \centering
        \includegraphics[width=0.99\textwidth]{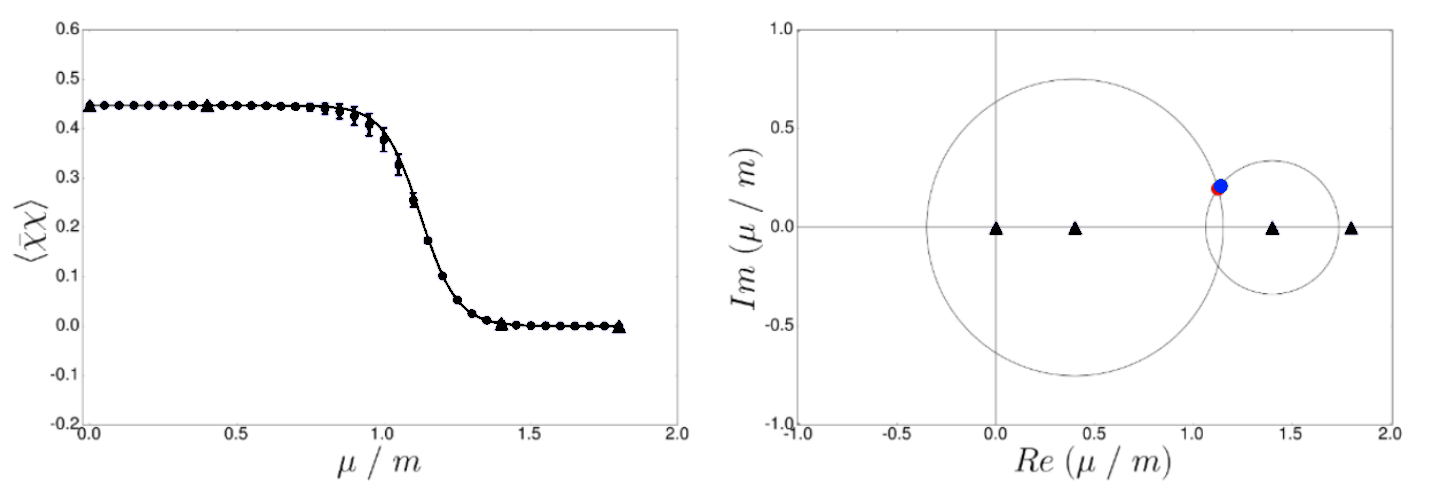} %
    \caption{Results for the one-dimensional Thirring model}
    \label{fig:thirring_res}
\end{figure}

The left picture of fig. \ref{fig:thirring_res} shows the results obtained for  the scalar condensate
from the  Pad\'e interpolation of the Taylor coefficients calculated at $\frac{\mu}{m} = 0.0, 0.4, 1.4$ and $1.8$
respectively up to order $0, 2, 5$ and $0$. The numerical results (error bars)
are in good agreement with the analytical result (solid line). The right picture shows how the stable, uncancelled root of the
denominator of the Pad\'e approximant (blue point) matches very well the singularity of the analytical solution (red point) on the
complex $\frac{\mu}{m}$ plane. In ref. \cite{DiRenzo:2021kcw} we have also repeated this analysis towards the continuum limit.

\FloatBarrier

As a second application we have considered the simple version of heavy-dense QCD \cite{philipsen1,philipsen2} that we had initially investigated in ref. \cite{Zambello:2018ibq}.
We have used the parameters $N_s^3 = 2^3, N_t = 116$ and $k = 0.0000887$. We have calculate Taylor expansions in term of the dimensionless
parameter $h_1 = (2 k e^{\hat{\mu}})^{N_t} = e^{-\frac{\mu - m}{T}}$ (actually we expanded in term of $h_1$ for $\mu < m$ and in term of $h_1^{-1}$ for $\mu > m$).
This is the natural expansion term, as the chemical potential $\mu$ enters the lattice action only through $h_1$.
As expansion points we chose $\frac{\mu}{m} = 0.9995, 1.0000, 1.0005$. From a semiclassical calculations we have
concluded that at these chemical potentials the contributions from all but the fundamental thimble are depressed \cite{DiRenzo:2020cgp}.

\begin{figure}[htp]
    \centering
        \includegraphics[width=0.99\textwidth]{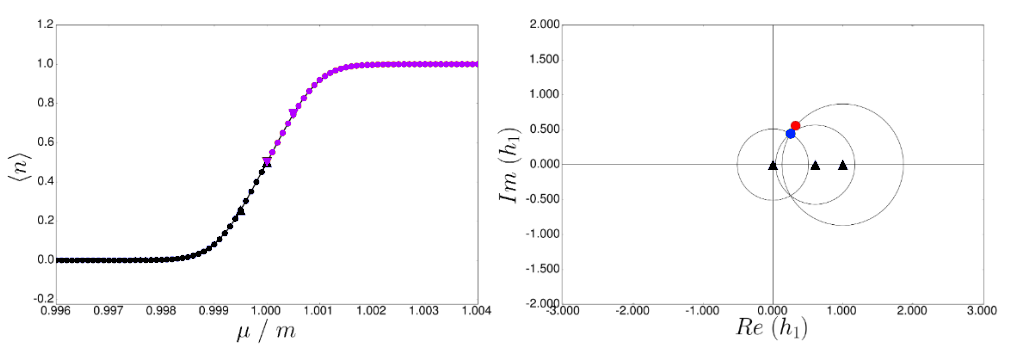} %
    \caption{Result for heavy-dense QCD}
    \label{fig:hdqcd_res}
\end{figure}

The results for the number density from Pad\'e are shown in the left picture of fig. \ref{fig:hdqcd_res},
using different colors for the $\mu < m$ and $\mu > m$ branches.
The Pad\'e interpolation for the left branch has been calculated using as inputs the Taylor coefficients up to order $2$ and $1$ 
respectively at $\frac{\mu}{m} = 0.9995$ and $1.0000$. We also added an extra constraint for the observable and its first
derivative at $\frac{\mu}{m} = 0$. Similarly for the right branch we used the Taylor coefficients up to order $1$ and $2$
at $\frac{\mu}{m} = 1.0000$ and $1.0005$, adding as extra constraint the asymptotic values of the observable and its first derivative
at large chemical potentials. Also in this second application the numerical results (error bars) are in agreement with
the analytical solution (solid line). The right picture shows that also in this case the uncancelled root of the denominator of the
approximant (blue point) matches fairly well the singularity of the observable (red point) in the complex $h_1$ plane.

\FloatBarrier

\section{Applications beyond thimble regularization}
As we've seen in the examples we've examined in the previous section, merging different Taylor series by Pad\'e not only improves
the convergence properties of the series, but it also returns information on the singularities structure of the observables. 
This is something that can be valuable in lattice QCD calculations. One might be able to extract hints on the location of
critical points in the QCD phase diagram. We stress that the strategy we have put in place is applicable with
any calculation method that gives access to Taylor expansions around multiple points. A natural question is whether the strategy can be
fruitful when applied to lattice calculation at imaginary chemical potentials (where there is no sign problem).

\begin{figure}[htp]
    \centering
        \includegraphics[width=0.48\textwidth]{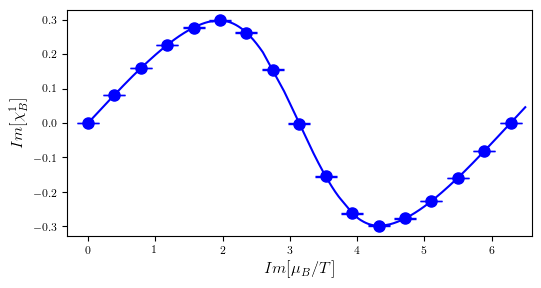} %
        \includegraphics[width=0.48\textwidth]{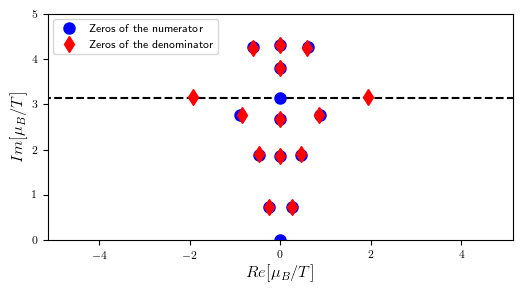} %
    \caption{Results from lattice QCD calculations at imaginary $\mu_B$}
    \label{fig:immu_res}
\end{figure}

We have been exploring this line of research in collaboration with the Bielefeld group \cite{Schmidt:2021pey,newpaper}.
Indeed we were able to locate a few singularities in the complex $\mu_B$ plan. For instance the left picture of
fig. \ref{fig:immu_res} shows the approximant we have built for the imaginary part of the baryon number density.
The approximant was  built from the lattice data obtained at $O(10)$ imaginary baryon chemical potentials
for the number density and its $1st$ and $2nd$ order derivatives. The right picture of fig. \ref{fig:immu_res} shows how two
roots of the denominator (red squares) survive the cancellation with roots of the numerator (blue points).
The location of these roots is stable with respect to variations of the inputs we use for Pad\'e and it signals
the presence of a genuine singularity.

\FloatBarrier

\FloatBarrier

\section{Conclusions}
One of the main issues that prevents the application of thimble regularization to realistic theories
is the calculation of the relative weights of the thimble contributions.

We argue that one can exploit the thimble structure of a theory to bypass the need for such calculations.
If one can find multiple \textit{good} points in the parameter space of the theory where only one thimble contribution
matters, one can calculate different Taylor expansions around these points. The Taylor coefficients
can be calculated by one-thimble simulations. The Taylor series can then be bridged by Pad\'e approximants
in order to reach \textit{bad} regions where more than one relevant thimble enter the thimble decomposition.
Moreover by studying the poles of the Pad\'e approximants one is able to extract information about the singularities structure
of the theory.

We also argue that this strategy has potential applications beyond thimble regularization itself. Indeed, any calculation method that
is good enough to give access to multiple expansion points can be used to build multi-point Pad\'e approximants. 
Specifically, the strategy can be useful to hunt for singularities in the phase diagram of QCD by merging Taylor series calculated
at imaginary chemical potentials.

\section{Acknowledgments}
This work was supported by the European Union Horizon 2020 research and innovation
programme under the Marie Sklodowska-Curie grant agreement No 813942 (EuroPLEx) 
and by the I.N.F.N. under the research project i.s. QCDLAT. This research used
computing resources made available (i) by CINECA on Marconi and Marconi 100 under
both the I.N.F.N-CINECA agreement and the ISCRA C program, (ii) by the University of Parma
on its HPC (High Performance Computing) facility, (iii) by the Gauss Centre for Supercomputing on
the Juwels GPU nodes at the J\"ulich Supercomputing Centre and (iv) by the Bielefeld University
on the Bielefeld GPU-Cluster.

\end{document}